\documentclass[twocolumn,showpacs,prl,amsmath,amssymb,floatfix]{revtex4}
\usepackage{amsmath,amssymb}
\usepackage{bm}
\usepackage{epsfig}
\usepackage{psfrag}

\newcommand{\bd}{\bm}

\begin{document}

\title{
Probing Anomalous Longitudinal Fluctuations
of the Interacting Bose Gas via Bose-Einstein Condensation of Magnons
}

\author{
Andreas Kreisel, Nils Hasselmann, and Peter Kopietz}
  
\affiliation{Institut f\"{u}r Theoretische Physik, Universit\"{a}t
  Frankfurt,  Max-von-Laue Strasse 1, 60438 Frankfurt, Germany}

\date{October 26, 2006}

 \begin{abstract}

The emergence of a finite staggered magnetization
in quantum Heisenberg antiferromagnets subject to a uniform magnetic field
can be viewed as Bose-Einstein condensation of magnons.
Using non-perturbative results for the 
infrared behavior
of the interacting Bose gas, we present exact results for
the staggered spin-spin correlation functions of quantum antiferromagnets
in a magnetic field at zero temperature. 
In particular, we show that  in dimensions $1 < D \leq 3$
the longitudinal
dynamic structure factor $S_{\parallel} ( {\bd{q}} , \omega )$
describing  staggered spin fluctuations 
in the direction of the staggered magnetization
exhibits a critical continuum whose weight can be controlled experimentally
by varying the magnetic field.

\end{abstract}


\maketitle

Because
magnons in ordered Heisenberg magnets are
bosonic quasi-particles, 
it is natural to expect that 
under certain conditions 
these systems
can be used to study general properties of
interacting Bose gases, such as the phenomenon of Bose-Einstein condensation (BEC).
In fact,  the renewed interest in BEC in recent years
 has also led to considerable 
activity in the field of  magnon BEC \cite{Nikuni00,Radu05}, which we 
abbreviate MBEC below.

Experimentally, MBEC
has been observed in two different classes of systems.
The first are  spin-singlet systems \cite{Nikuni00} such as
TlCuCl$_3$ or Haldane gap spin chains, 
where  MBEC corresponds to the collapse of the singlet-triplet gap
at a critical magnetic field and the emergence of long-range magnetic order for
larger fields.
The second class of systems, which we shall consider in this work,
are easy-plane quantum antiferromagnets (QAFs)
such as Cs$_2$CuCl$_4$ subject to a magnetic field $h$ perpendicular to the 
easy plane \cite{Radu05}.
If $h$ exceeds some critical field  $h_c$, the ground state is a saturated ferromagnet.
For $h < h_c$ the
$U(1)$-symmetry of the Hamiltonian is  spontaneously broken
and antiferromagnetic long-range order emerges.
At zero temperature, the boson density vanishes at the phase transition 
so  that many-body interactions  are not relevant at the zero temperature
quantum critical point. However, the finite temperature transition
belongs to the same universality class 
as BEC in the interacting Bose gas. 
One advantage of studying BEC in magnetic systems is that
 the magnetic analogon $h_c-h$
of the chemical potential can easily be controlled experimentally.
Another advantage, which has attracted little attention to date, is that
in MBEC
the phase of the condensate has a direct physical interpretation
as the orientation of the staggered magnetization within the easy 
plane.

The concept of MBEC has been formulated theoretically many years 
ago \cite{Matsubara56,Batyev84,Batyev85,Affleck90},
where the focus was mainly 
on the nature of the quantum phase transition. 
However, even away from the critical point, perturbation theory for the
interacting Bose gas is plagued by divergences originating from the 
gapless dispersion of the Goldstone mode,
giving rise to non-analytic  behavior of some  boson correlation  
functions~\cite{Castellani97,Weichman88,Giorgini92}. Yet, in all physical observables
of the Bose gas
these divergences cancel, so that they seem to be little more
than a mathematical curiosity. Here, we point out that, in contrast 
to the Bose gas, in MBEC the anomalous behavior arising from the divergences 
can be directly observed experimentally.
We shall exploit recent renormalization group results~\cite{Castellani97} to
obtain the infrared behavior of the staggered spin-spin correlation functions 
of QAFs in a uniform magnetic field
at zero temperature. 
In particular, we shall show that in dimensions $1<D \leq 3$ the 
longitudinal part $S_{\parallel} ( {\bd{q}} , \omega )$ 
of the dynamic structure factor (corresponding to spin-fluctuations parallel
to the staggered magnetization)
exhibits a critical continuum
for small wave-vectors $| {\bd{q}} |$ and frequencies $\omega$,
which can be measured via neutron scattering.

We  start from the Hamiltonian of the
spin-$S$ QAF on a $D$-dimensional
hypercubic lattice with $N$ sites and lattice spacing $a$ in a uniform magnetic field
in $z$-direction,
 \begin{equation}
  {H} = \frac{1}{2} \sum_{ ij} J_{ ij} {\bd{S}}_i \cdot {\bd{S}}_j  - h \sum_i S^{z}_i
 \; ,
 \label{eq:hamiltonian}
 \end{equation}
where the sums are over all sites of the lattice, and ${\bd{S}}_i$ 
are spin operators with ${\bd{S}}_i^2 = S (S+1)$. We assume nearest 
neighbor coupling, i.e.,
 $J_{ ij} = J > 0$ if 
$i$ and $j$ are nearest neighbors, and
$J_{ij}=0$ otherwise.
For sufficiently large $h$, the ground state of Eq.~(\ref{eq:hamiltonian}) is a saturated
ferromagnet with magnetization parallel to the field.
To obtain the low-lying magnon excitations, we express the spin operators
in terms of canonical boson operators $b_i$ using the
Holstein-Primakoff transformation, $S^{-}_i = \sqrt{2S}  b_i^{\dagger} 
[ 1 - n_i /(2S) ]^{1/2} = (S^{+}_i )^{\dagger}$, $S^z_i = S - n_i$,
where $n_i = b^{\dagger}_i b_i$. Expanding the square roots and
neglecting terms involving six and more
boson operators we obtain ${H} \approx E_0 + H_2 + H_4$.
The constant part is
$E_0 = N [ \tilde{J}_0 S^2/2 - hS ]$, where $\tilde{J}_0 = \tilde{J}_{ \bd{k} =0}$ with
$\tilde{J}_{\bd{k}} = \sum_j e^{ i {\bd{k}} \cdot {\bd{r}}_j }J_{ ij}$. 
The quadratic  and the quartic contributions to the Hamiltonian are
 \begin{eqnarray}
 H_2 & = & - \frac{S}{2} \sum_{ij} J_{ij} [ n_i + n_j - b^{\dagger}_i b_j - b^{\dagger}_j b_i ] + h \sum_i n_i
  , \; \; 
 \label{eqw:H2def}
 \\
 H_4 & = & \frac{1}{4} \sum_{ij} J_{ij} [ 2 n_i n_j -
  b_i^{\dagger} n_i b_j 
 -  b_j^{\dagger} n_i b_i      ]
  \; .
 \label{eq:H4def}
 \end{eqnarray}
\vspace{-.2cm}

\vspace{-.2cm}
Anticipating antiferromagnetic symmetry breaking, it is
convenient to
define the Fourier transform in the reduced
(antiferromagnetic) Brillouin zone with
two branches of magnon operators $b_{ \bd{k} \sigma}$ labelled by 
momenta ${\bd{k}}$ in the reduced zone and
$\sigma = \pm $,
 \begin{equation}
 b_{\bd{k}  \sigma} = N^{-1/2}  \bigl[
 \sum_{ i \in A}   e^{ - i {\bd{k}} \cdot {\bd{r}}_i } b_i
 +    \sigma   \sum_{  i \in B}  e^{ - i {\bd{k}} \cdot {\bd{r}}_i } b_i \bigr]
\; .
 \label{eq:FT}
 \end{equation}
Here $i \in A/B$ denotes a summation over the sites of only one of the sublattices $A$ or $B$.
The quadratic magnon Hamiltonian is then diagonal,
$H_2 = \sum_{ \bd{k} \sigma} ( \epsilon_{ \bd{k}  \sigma} - \mu ) b^{\dagger}_{\bd{k} \sigma} b_{ \bd{k} \sigma}$,
with magnon dispersion
$\epsilon_{ \bd{k}  \sigma} =   ( \tilde{J}_0   + \sigma \tilde{J}_{\bd{k}} )S$ and
$\mu = h_c  - h $, where $h_c =2 \tilde{J}_0 S$.
For small wave-vectors
 $\epsilon_{ \bd{k} , -} \approx {\bd{k}^2}/(2m) $, 
with effective mass $ m  = (2 JS a^2)^{-1}$, while
the symmetric mode $\epsilon_{ \bd{k} , +}$ is gapped,
$\epsilon_{ \bd{k}=0 , +} = h_c$, so that 
$\epsilon_{ \bd{k}=0 , +} - \mu = h$. 
Note that the gapless mode $\sigma =-1$ describes staggered spin fluctuations.

Since in this work we are interested in the infrared behavior of 
correlation functions
at energy scales smaller than $h$, we shall ignore the gapped mode.
To derive the interaction part of the  Hamiltonian 
of the gapless magnons,
we substitute Eq.~(\ref{eq:FT}) into Eq.~(\ref{eq:H4def})
and neglect all terms
involving the operators $b_{\bd{k}, +}$. 
Defining the field operators $\hat{\psi}_{ \bd{k} } = V^{1/2} b_{ \bd{k} , - } $,
where $V = N a^D$ is the volume of the system,
 we obtain for $N \rightarrow \infty$,
 \begin{eqnarray}
 H & = & E_0 + \int_{ \bd{k} } ( \frac{ \bd{k}^2}{2m} - \mu )  \hat{\psi}_{\bd{k}}^{\dagger}
  \hat{\psi}_{\bd{k}}
 \nonumber
 \\
 & + & \frac{1}{2} 
\int_{ \bd{q} }  \int_{ \bd{k} }  \int_{ \bd{k}^{\prime} } 
 U_{\bd{q}} 
  \hat{\psi}_{\bd{k} + {\bd{q}} }^{\dagger} \hat{\psi}_{\bd{k}^{\prime} - {\bd{q}} }^{\dagger}
 \hat{\psi}_{\bd{k}^{\prime}  } \hat{\psi}_{\bd{k} }
 \; ,
 \label{eq:Bosehamiltonian}
 \end{eqnarray}
\vspace*{-.3cm}

\noindent where $\int_{\bd{k}} = \int d^D k / ( 2 \pi )^D$ and the effective
interaction is $U_{\bd{q}}  = \Theta ( \Lambda_0 -  | {\bd{q}} | )  \chi_0^{-1}$.
Here $\Lambda_0 \approx a^{-1}$ is an ultraviolet cutoff and
$\chi_0 = (2 \tilde{J}_0 a^D )^{-1}$ is the leading large-$S$ result for
the uniform transverse susceptibility in $x$-direction, which
is perpendicular to the staggered magnetization and to the
magnetic field.
This completes our mapping of the spin system onto
an interacting Bose gas.

\begin{figure}[tb]
  \centering
  \psfrag{t}{$\theta$}
  \psfrag{h}{$\bd{h}$}
  \psfrag{ms}{${\bd{M}}_s$}
  \epsfig{file=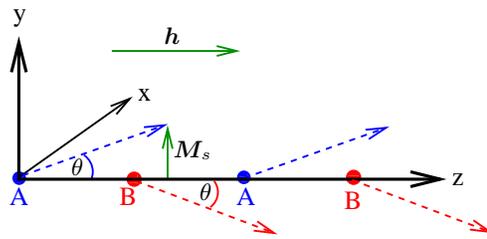,width=65mm}
  \vspace{-4mm}
  \caption{%
(Color online) Spin configuration  in the ground state of a QAF in a uniform magnetic field
for $h < h_c$. The local moments  $\langle {\bd{S}}_i \rangle$ are represented by
dashed arrows.
We choose the coordinate system such that the magnetic field ${\bd{h}}$   points
in the $z$-direction and  
the staggered magnetization $\bd{M}_s$ points in the $y$-direction. 
The sublattices are labelled A and B.}
  \label{fig:tilt}
\end{figure}

For $ \mu > 0$ the field operator
$\hat{\psi}_{ \bd{k} =0}$ acquires a vacuum expectation value, corresponding
to the emergence of long-range antiferromagnetic order. 
The spin configuration in the ground state is then canted, 
as shown in Fig.~\ref{fig:tilt}.
One should keep in mind, however, that Eq.~(\ref{eq:Bosehamiltonian})
is accurate
only for small $\theta$, since otherwise the Bose
gas is no longer dilute such that 
a three-body interaction must be added.
However, the qualitative behavior of correlation functions
is largely determined by
symmetry \cite{Castellani97}, so that
our results derived below are expected to  
remain valid also for larger canting angles. 
Of course, if $\theta$ is not small it is
better to quantize the spin operators 
in a suitably defined tilted basis \cite{footnote3,Hasselmann06b}.
For simplicity, we shall focus here on the regime 
$\theta \ll 1$ where our approach
based on the Holstein-Primakoff transformation with quantization axis along the
direction of the magnetic field is thus quantitatively
accurate for large $S$. Alternatively, we could represent
the spin operators in terms of projected 
bosons~\cite{Batyev84,Batyev85,footnote1} which account
for all $1/S$ corrections at small $\theta$. However,
in the spirit of Ref.~[\onlinecite{Chakravarty88}], 
we take the finite corrections due to higher orders in $1/S$ 
implicitly into account
by expressing our final result in terms of the
true spin-wave velocity $c$ and the spin susceptibility $\chi$,
which can be obtained from experiments.
 
To leading order in $1/S$,
the classical tilt angle $\theta$ 
is easily obtained 
by substituting $\hat{\psi}_{\bd{k}}  = ( 2 \pi )^D \delta ( \bd{k} ) \psi_0 + \Delta
 \hat{\psi}_{\bd{k}}$ in Eq.~(\ref{eq:Bosehamiltonian})
and demanding that the coefficient of the term
linear in $\Delta \hat{\psi}_{\bd{k}}$ vanishes.
This yields the well known \cite{Popov87} relation $| \psi_0 |^2 U_0 = \mu$.
To translate this into a condition for $\theta$,
we note from Fig.~\ref{fig:tilt} that for large $S$ 
the tilt angle is related to the staggered magnetization
$M_s = V^{-1} \sum_i \zeta_i \langle S^y_i \rangle$ via
$\theta \approx M_s / s$. 
Here  $s = S / a^D$ is the spin-density of the 
saturated magnetic state,
and
$\zeta_i =1$ for $i \in A$ and
$\zeta_i =-1$ for $i \in B$.
Since
$M^2_s = 2s | \psi_0 |^2$ to leading order in spin-wave theory, 
we obtain for large $S$
 \begin{equation}
  \theta^2  \approx M_s^2 / s^2 \approx 2 \rho_0 / s  =  2 (1 - h/h_c) 
 \; ,
 \label{eq:thetares}
 \end{equation}
where 
$\rho_0 = | \psi_0 |^2$ is the condensate density. 
Eq.~(\ref{eq:thetares}) agrees with
the small-$\theta$ expansion of the classical result
$\cos \theta = h / h_c$ \cite{Zhitomirsky98}.
To obtain the magnon spectrum for large $S$, 
we neglect all terms higher than quadratic in the 
$\Delta \hat{\psi}_{\bd{k}}$. 
After a Bogoliubov transformation, one finds
that for $h<h_c$ the magnon spectrum is determined by
$\epsilon_{\bd{k}}^2=\big(\frac{{\bd k}^2}{2m}\big)^2 + c_0^2 {\bd k}^2$
which for small $\bd{k}$ yields
$\epsilon_{\bd{k}} = c_0  | {\bd{k}} | + {\cal O}(\bd{k}^3)$, with 
the magnetic field dependent
spin-wave velocity $c_0 = \sqrt{\mu / m } =  2 \sqrt{D} JS a  \theta$.

We now focus on the spin-wave interactions, which are 
relevant in $D \leq 3$.
The substitution $\hat{\psi}_{\bd{k}}  = ( 2 \pi )^D \delta ( \bd{k} ) \psi_0 + \Delta
 \hat{\psi}_{\bd{k}}$ generates various terms cubic and quartic
in $\Delta
 \hat{\psi}_{\bd{k}}$
which are rather tedious to work with. 
A perturbative treatment of these terms is plagued 
by infrared divergences. However,
Ward-identities associated with the $U(1)$-symmetry 
of the Hamiltonian (\ref{eq:Bosehamiltonian}) can be used to obtain the
infrared behavior of some correlation functions without resorting to
perturbation theory.
As pointed out in Ref.~\cite{Castellani97},
the constraints imposed on the correlation functions
by the Ward identities are more transparent
if one expresses $\hat{\psi}_{\bd{k}}$ in terms of two real fields,
corresponding to transverse and longitudinal fluctuations.
Recently we have proposed a similar parameterization of the
spin-wave expansion in  QAFs \cite{Hasselmann06,Hasselmann06b}, which
amounts to expressing  the
field operator $\hat{\psi}_{\bd{k}}$ in terms of two canonically conjugate
hermitian operators
$\hat{\Pi}_{\bd{k}}$ and $\hat{\Phi}_{\bd{k}}$ as 
follows~\cite{footnote2}, 
 \begin{equation}
 \hat{\psi}_{\bd{k}} = \rho_0^{1/2} \hat{\Pi}_{\bd{k}} 
+ i ( 4 \rho_0 )^{-1/2}  \hat{\Phi}_{\bd{k}}
 \; .
 \label{eq:realfield}
 \end{equation}
Both fields represent staggered spin fluctuations in the
directions perpendicular to the magnetic field,
where
$\hat{\Pi}_{\bd{k}}$ is transverse and
$\hat{\Phi}_{\bd{k}}$ longitudinal with respect to the
staggered magnetization.
More precisely, in the coordinate system
shown in Fig.~\ref{fig:tilt} we have for $S\to \infty$,
 \begin{equation}
   \hat{\Pi}_{\bd{k}}   \approx 
  \frac{1}{M_s} \sum_i \zeta_i e^{ -i {\bd{k}} \cdot {\bd{r}}_i } S^{x}_i
 \; \;  , \; \; 
 \hat{\Phi}_{\bd{k}}   \approx \frac{ M_s}{s} 
   \sum_i   \zeta_i  e^{ -i {\bd{k}}   \cdot {\bd{r}}_i }  S^{y}_i 
 \label{eq:PiSz}
 \; .
 \end{equation}
This equation becomes quantitatively accurate
even for  finite $S$
in the dilute limit $h\to h_c$ 
which can be shown using
projected boson operators \cite{footnote1}.  
The condensed phase corresponds to antiferromagnetic order,
$\langle \hat{\Phi}_{\bd{k}} \rangle  
= (2 \pi )^D \delta ( \bd{k} )  \phi_0$, where
$\phi_0 =  M_s^2 /s$ for large $S$.
Note that the total density of the Bose gas corresponds in the underlying
spin system to $\rho =  s  - V^{-1} \sum_i \langle S^z_i \rangle$, which vanishes for 
$h \geq h_c$ and  satisfies $ \rho \ll s$ if
$h$ is only slightly smaller than $h_c$.
Corrections
to Eq.~(\ref{eq:PiSz}) would be associated with terms which are
of linear or higher order in the boson density and become
negligible for $h\to h_c$.

The  infrared behavior of correlation functions is most conveniently 
derived within a functional integral approach \cite{Popov87,Castellani97}.
The system is then described in terms of an action
$S [ \Pi, \Phi ]$, which is a functional of
$c$-number fields depending on imaginary time $\tau$. Formally,
these fields can be obtained from the field operators by substituting
 $ \hat{\Pi}_{\bd{k}} \rightarrow  \Pi_{ \bd{k}} ( \tau )    $  and
$      \hat{\Phi}_{\bd{k}}  \rightarrow 
( 2 \pi )^D \delta ( \bd{k} ) \phi_0 + \Phi_{\bd{k}} ( \tau )$.
Introducing the Fourier transforms in frequency space,
$\Pi_K = \int d \tau e^{ i \omega \tau } \Pi_{\bd{k}} ( \tau )$ and
$\Phi_K = \int d \tau e^{ i \omega \tau } \Phi_{\bd{k}} ( \tau )$, 
the Gaussian part of the action can be written as
\begin{eqnarray}
 S_0 [ \Pi , \Phi ] & = & \frac{1}{2} \int_K
 \Bigr[  \chi_0   c_0^2 \bd{k}^2 \Pi_{ -K } \Pi_K + \chi_0^{-1} \Phi_{-K} \Phi_K
  \nonumber
 \\
 & & + \omega  ( \Pi_{ -K} \Phi_K - \Phi_{-K} \Pi_K ) \Bigr]
 \; .
 \label{eq:S2res}
 \end{eqnarray}
Here $ K = ( {\bd{k}} , i \omega )$ is a collective label for momenta and frequencies,
and $\int_K = \int \frac{d \omega}{ 2 \pi }
 \int_{\bd{k}}$.
The correlation functions are within Gaussian approximation given by
 \begin{subequations}
 \begin{eqnarray}
 \langle \Pi_K \Pi_{K^{\prime}} \rangle_0 & = & \delta_{ K , -K^{\prime}} 
 \frac{ \chi_0^{-1}  }{   \omega^2 + c_0^2 {\bd{k}}^2  }
 \; ,
 \label{eq:PiPicorr}
 \\
 \langle \Pi_K \Phi_{K^{\prime}} \rangle_0 & = & 
\delta_{ K , -K^{\prime}}  
 \frac{  \omega    }{   \omega^2 + c_0^2 {\bd{k}}^2  }
 \; ,
\label{eq:PiPhicorr}
 \\
 \langle \Phi_K \Phi_{K^{\prime}} \rangle_0 & = & \delta_{ K , -K^{\prime}}  
 \frac{  \chi_0 c_0^2 {\bd{k}}^2    }{   \omega^2 + c_0^2 {\bd{k}}^2  }
 \; ,
\label{eq:PhiPhicorr}
 \end{eqnarray}
 \end{subequations}
where 
$\delta_{ K , - K^{\prime} } = (2 \pi )^{D+1} \delta ( \omega + \omega^{\prime} )
\delta ( {\bd{k}} + {\bd{k}}^{\prime} )$.
It turns out that the Gaussian approximation is qualitatively correct 
for the transverse correlation function
$ \langle \Pi_K \Pi_{K^{\prime}} \rangle$
and for the mixed correlation function
$ \langle \Pi_K \Phi_{K^{\prime}} \rangle$, where $\langle \dots\rangle$ 
denotes
the full thermal average.
Using the results of Ref.~[\onlinecite{Castellani97}],
we obtain for the true infrared behavior of these correlation functions
\vspace{-.1cm}
  \begin{subequations}
 \begin{eqnarray}
 \langle \Pi_K \Pi_{K^{\prime}} \rangle & = & \delta_{ K , -K^{\prime}} 
  \frac{  \chi^{-1} }{   \omega^2 + c^2 {\bd{k}}^2  }
 \; ,
 \label{eq:PiPicorrtrue}
 \\
 \langle \Pi_K \Phi_{K^{\prime}} \rangle & = & 
\delta_{ K , -K^{\prime}}
 \frac{ Z_{\parallel}   \omega    }{   \omega^2 + c^2 {\bd{k}}^2  }
 \; ,
\label{eq:PiPhicorrtrue}
 \end{eqnarray}
 \end{subequations}
where $c$ is the renormalized spin-wave velocity,
$\chi = M_s^2 \rho /  (2 s m c^2 \rho_0 )  = \chi_0 
 Z_{\rho}/ Z_c^2$  is the renormalized
transverse susceptibility, and
$\rho_0$ and $\rho$ are the true condensate density and total density.
For convenience we have introduced the 
dimensionless factors $Z_\rho = \rho / \rho_0$, $Z_c = c / c_0$, and
 $Z_{\parallel}  =  (mc^2 /{\rho})    d \rho_0 /d \mu $, all of which approach unity
in the classical limit $S \rightarrow \infty$.
We emphasize that
in dimensions $D \leq 3$ 
some Feynman diagrams  contributing to 
the above correlation functions are infrared divergent.
However, the Ward identities \cite{Castellani97} guarantee that
all divergences cancel, so that the only difference
between the exact results (\ref{eq:PiPicorrtrue}, \ref{eq:PiPhicorrtrue}) and the
corresponding approximate
expressions (\ref{eq:PiPicorr}, \ref{eq:PiPhicorr})
are finite renormalization factors. The exact results (\ref{eq:PiPicorrtrue}, \ref{eq:PiPhicorrtrue}) depend only on parameters which can be expressed in terms of 
thermodynamic derivatives.

The important point is now that the Ward identities do not impose similar constraints on the longitudinal correlation
function $ \langle \Phi_K \Phi_{K^{\prime}} \rangle$, which 
in  $D \leq 3$ is dominated by a
non-analytic term arising from the re-summation of the leading
infrared divergences.  Using the results derived in  
Ref.~[\onlinecite{Castellani97}]
we obtain
\vspace{-.4cm}
 \begin{eqnarray}
  \langle \Phi_K \Phi_{K^{\prime}} \rangle & = &
\delta_{ K , -K^{\prime}}   \chi \biggl[   
 - Z_{\parallel}^2 \frac{   \omega^2  }{   \omega^2 + c^2 {\bd{k}}^2  }
 \nonumber
 \\
 &    &  \hspace{-23mm} +   K_{ D+1}   \frac{  ( mc)^3}{  Z_{\rho}^3 \rho_0}
 \Biggl\{
 \begin{array}{ll}
 \ln [  \frac{ (m c)^2  }{  \omega^2/ c^2 + {\bd{k}}^2  } ] 
 & \mbox{, $ D =3 $ }
 \\
 \frac{2}{ 3 - D }   [ \frac{\omega^2}{ c^2} + {\bd{k}}^2  ]^{  \frac{D - 3 }{2} } 
 & \mbox{, $ D <3 $ }
\end{array}
 \Biggr]
  \; ,
 \label{eq:PhiPhicorrtrue}
\end{eqnarray}
where $K_D = 2^{1-D} \pi^{- D/2}/ \Gamma (D/2)$.
The non-analytic contribution 
to the longitudinal correlation function
of the interacting Bose gas has first been discussed
by Weichman~\cite{Weichman88}, 
and later in Ref.~[\onlinecite{Giorgini92}]. The field $\phi_K$ 
has thus a finite anomalous dimension and its effective action cannot be 
approximated by a Gaussian. 
This reflects the general behavior of systems 
with  broken continuous symmetries where Goldstone modes
lead to 
 anomalous longitudinal 
fluctuations~\cite{Patashinskii73,Chakravarty91,
Giorgini99,Sachdev99b,Zwerger04}.
Note that
the  magnetic field dependent microscopic momentum scale $mc$ is
for large $S$
approximately
$mc \approx \sqrt{D} \theta /a$.
Using a generalized Ginzburg criterium~\cite{Castellani97},
we  find that the non-analytic corrections
in Eq.~(\ref{eq:PhiPhicorrtrue}) become important 
for $|\bd{k}|\lesssim k_G$, where
$k_G\approx m c \, [(m c)^D/ \rho_0]^{\frac{1}{3-D}}$ for $D<3$, and 
$k_G\approx m c \exp[- \rho_0/(m c)^3]$  for $D=3$.

Even though we have derived 
Eqs.~(\ref{eq:PiPicorrtrue}, \ref{eq:PiPhicorrtrue}) and  (\ref{eq:PhiPhicorrtrue}) only for small canting angle $\theta$,
they remain valid even for moderate values
of $\theta$, as long as three-body interactions are negligible.
Using Eqs.~(\ref{eq:PiSz},\ref{eq:PhiPhicorrtrue}), we find for the 
longitudinal staggered structure factor for $1<D \leq 3$
and $ 0< \omega /c \lesssim k_G$,
 \begin{eqnarray}
 S_{\parallel} ( {\bd{k}} , \omega ) & = &   \frac{\chi \, s^2}{M_s^2}
 \Biggl[
  \frac{Z_{\parallel}^2}{2} 
     c | {\bd{k}} |
  \delta ( \omega - c | {\bd{k}} |)
 \nonumber
 \\
 & + & C_D
 \frac{ (mc)^3}{Z_{\rho}^3 \rho_0}
  \frac{  \Theta ( \omega - c | \bd{k} | )  }{ 
  ( \omega^2 /c^2  - {\bd{k}}^2  )^{  \frac{3-D}{2} } }
 \Biggr]
 \; ,
 \label{eq:dynres}
\end{eqnarray}
where $C_D =  K_{D+1} [\pi ( 3 -D)/2]^{-1}
  \sin [  \pi ( 3-D )/2 ]$.
In particular, $C_3 = K_4 = ( 8 \pi^2 )^{-1}$ and
$C_2  = \pi^{-3}$. 
Eq.~(\ref{eq:dynres}) is the main result of this work. 
For small $ | {\bd{k}} |\lesssim k_G$
the critical continuum represented by the 
last term in Eq.~(\ref{eq:dynres}) 
carries most of the spectral weight:
Denoting by $I_c$ the contribution from the continuum part
 in Eq.~(\ref{eq:dynres}) to the energy integrated spectral
weight, and by $I_\delta$ the corresponding contribution
due to the $\delta$-function, we find
$I_c/I_\delta\propto (k_G/|\bd{k}|) \ln(k_G/|\bd{k}|)$
for $D=2$ 
and $I_c/I_\delta\propto (k_G/|\bd{k}|) 
(mc)^3/\rho_0$
for $D=3$. 
In both cases the continuum part dominates for  $|\bd{k}|\lesssim k_G$.
However, in three dimensions $k_G\propto \theta \exp[-S/(6 \sqrt{3} \theta)]$
is exponentially small
for $h\approx h_c$ so that the region $|\bd{k}|\lesssim k_G$ is difficult
to resolve experimentally.
On the other hand, in $D=2$ we have 
$k_G\approx 4 \sqrt{2} \theta/S a$ so that in this case the 
critical continuum should be accessible via neutron scattering
at scales $| \bd{k} | \lesssim k_G$.

In summary, we have shown that in dimensions $1<D \leq 3$
the longitudinal dynamic structure factor
of QAFs subject to  a uniform magnetic field 
exhibits a field-dependent critical continuum
as described by Eq.~(\ref{eq:dynres}) which
is closely related
to the anomalously large longitudinal fluctuations in the condensed phase
of the interacting Bose gas.
While in the latter case these fluctuations cannot be directly measured,
the magnetic realization of BEC offers 
a direct experimental access to these fluctuations.
Both in $D=2$ and $D=3$ the critical continuum dominates
the energy integrated longitudinal structure factor if
$ | \bd{k} |$ is
smaller than the Ginzburg scale $k_G$. 
At least in two-dimensional QAFs the critical continuum should
be observable in neutron scattering experiments.

This work was supported by the DFG via FOR 412.


\vspace*{-.5cm} 

\begin{thebibliography}{99}
%
\bibitem{Nikuni00}
T. Nikuni, M. Oshikawa, A. Oosawa, and H. Tanaka,
Phys. Rev. Lett. {\bf{84}}, 5868 (2000);
R. Coldea, D. A. Tennant, K. Habicht, P. Smeibidl, C. Woltzers, and Z. Tylczynski,
Phys. Rev. Lett. {\bf{88}}, 137203 (2002);
M. Matsumoto, B. Normand, T. M. Rice, and M. Sigrist,
Phys. Rev. Lett. {\bf{89}}, 077203 (2002);
T. M. Rice, Science {\bf{298}}, 760 (2002);
Ch. R\"{u}egg, N. Cavadini, A. Furrer, H. U. G\"{u}del, K. Kr\"{a}mer, H. Mutka, A.
Wildes, K. Habicht, and P. Vorderwisch, Nature {\bf{423}}, 62 (2003);
V. S. Zapf, D. Zocco, B. R. Hansen, M. Jaime, N. Harrison, C. D. Batista, 
M. Kenzelmann, C. Niedermayer, A. Lacerda, and A. Paduan-Filho,
Phys. Rev. Lett. {\bf{96}}, 077204 (2006).
%
\bibitem{Radu05}
T. Radu, H. Wilhelm, V. Yushankhai, D. Kovrizhin, R. Coldea, Z. Tylczynski, T.
L\"{u}hmann, and F. Steglich, Phys. Rev. Lett. {\bf{95}}, 127202 (2005).
%
%
\bibitem{Matsubara56}
T. Matsubara and H. Matsuda, Prog. Theor. Phys. {\bf{16}}, 569 (1956).
%
\bibitem{Batyev84}
E. G. Batyev and L. S. Braginskii, Zh. Eksp. Teor. Fiz. {\bf{87}}, 1361 (1984)
[Sov. Phys. JETP {\bf{60}}, 781 (1984)].
\bibitem{Batyev85}
 E.~G.~Batyev  Zh. Eksp. Teor. Fiz. 
{\bf{89}}, 308 (1985)
[Sov. Phys. JETP {\bf{62}}, 173 (1985)]; S. Gluzman, 
Z.  Phys. B {\bf 90}, 313 (1993).
%
\bibitem{Affleck90}
I. Affleck, Phys. Rev. B {\bf{41}}, 6697 (1990); {\it{ibid.}} {\bf{43}}, 3215 (1991).
%
\bibitem{Castellani97}
C. Castellani, C. Di Castro, F. Pistolesi, and G. C. Strinati, Phys. Rev. Lett.
{\bf{78}}, 1612 (1997);
F. Pistolesi, C. Castellani, C. Di Castro, and G. C. Strinati, 
Phys. Rev. B {\bf{69}}, 024513 (2004).
%
\bibitem{Weichman88}
P. B. Weichman,  Phys. Rev. B {\bf{38}}, 8739 (1988).
%
\bibitem{Giorgini92}
S. Giorgini, L. Pitaevskii, and S. Stringari, 
Phys. Rev. B {\bf{46}}, 6374 (1992).
%
\bibitem{footnote3}
Based on a self-consistent treatment of a subset of interaction
diagrams, a continuum in the dynamical structure factor was predicted
at large magnetic fields and {\em large} wavevectors in M.~E.~Zhitomirsky
and A.~L.~Chernyshev, Phys. Rev. Lett. {\bf{82}}, 4536 (1999). They
did however not include the quartic magnon interaction.
%
\bibitem{Hasselmann06b}
N. Hasselmann, F. Sch\"{u}tz, I. Spremo, and P. Kopietz, cond-mat/0511706,
(C. R. Chim., in press).
%
%
\bibitem{footnote1}
The usual hard core boson approach~\cite{Batyev84} for $S=1/2$ can 
be generalized for  arbitrary spin by setting 
$S_i^+\approx (2S)^{1/2}[1+(K-1)b_i^\dagger b_i]b_i$, 
with $K=(1-1/2S)^{1/2}$, see Ref.~[\onlinecite{Batyev85}]. 
To leading order in $1/S$,
this reproduces the interaction term
in Eq.~(\ref{eq:H4def}).
In the low-density limit $\theta \ll 1$ this projected boson
approach 
becomes exact. 
%
\bibitem{Chakravarty88}
S. Chakravarty, B. I. Halperin, and D. Nelson, Phys. Rev. Lett. {\bf{60}}, 1057 (1988); Phys. Rev. B {\bf{39}}, 2344 (1989).
%
\bibitem{Popov87}
V. N. Popov, {\it{Functional Integrals and Collective Excitations}},
(Cambridge University Press, Cambridge, 1987).
%
\bibitem{Zhitomirsky98}
M. E. Zhitomirsky and T. Nikuni, Phys. Rev. B {\bf{57}}, 5013 (1998).
%
\bibitem{Hasselmann06}
N. Hasselmann and P. Kopietz, Europhys. Lett. {\bf{74}}, 1067 (2006).
%
\bibitem{footnote2}
The normalization in Eq.~(\ref{eq:realfield}) is chosen to identify
$\hat{\Pi}_{\bd{k}}$ with the transverse field of
the non-linear 
$\sigma$-model~\cite{Hasselmann06,Hasselmann06b,Chakravarty88}.
The conjugate
field 
$\hat{\Phi}_{\bd{k}}$ is then fixed by the 
canonical
commutation relation
$ [ \hat{\Pi}_{\bd{k}} , \hat{\Phi}_{\bd{k}^{\prime}} ]
= i ( 2 \pi )^D \delta ( \bd{k} + \bd{k}^{\prime} )$.
%
\bibitem{Patashinskii73}
A. Z. Patashinskii and V. L. Pokrovskii, Zh. Eksp. Teor. Fiz. {\bf{64}},
1445 (1973) [Sov. Phys. JETP {\bf{37}}, 733 (1973)].
%
\bibitem{Chakravarty91}
S. Chakravarty, Phys. Rev. Lett. {\bf{66}}, 481 (1991).
%
\bibitem{Giorgini99}
S. Giorgini, L. P. Pitaevskii, and S. Stringari, Phys. Rev. Lett. {\bf{80}}, 5040 (1999).
%
\bibitem{Sachdev99b}
S. Sachdev, Phys. Rev. B {\bf{59}}, 14054 (1999).
%
\bibitem{Zwerger04}
W. Zwerger, Phys. Rev. Lett. {\bf{92}}, 027203 (2004).
%





\end{thebibliography}
\end{document}